\documentclass[journal,11pt]{article}

\usepackage{algorithm}
\usepackage{algorithmic}
\usepackage{authblk}
\usepackage{amsthm}
\usepackage{amsmath}
\usepackage{thmtools,thm-restate}
\usepackage{amssymb}
\usepackage{graphicx}
\usepackage{commath}
\usepackage{xcolor}
\usepackage{color}
\usepackage{url}
\usepackage{bm}
\usepackage{verbatim}
\usepackage{subcaption}
\usepackage{enumitem}

\numberwithin{equation}{section}

\newcommand{\R}{\mathbb{R}}

\newcommand{\n}{\hat{n}}

\newcommand{\N}{\mathcal{N}}

\newcommand{\gap}{\text{Gap}}

\begin{document}
%\title{A dynamic programming approach for signal detection}
\title{Detecting Non-overlapping Signals with Dynamic Programming}

\author[1]{Mordechai Roth}
\author[2]{Amichai Painsky}
\author[1]{Tamir Bendory}

\affil[1]{School of Electrical Engineering,	Tel Aviv University, Tel Aviv, Israel}
\affil[2]{Department of Indusrial Engineering, Tel Aviv University, Tel Aviv, Israel}

\maketitle

\begin{abstract}
	This paper studies the classical problem of detecting the locations of  signal occurrences  in a one-dimensional noisy measurement.
	Assuming the signal occurrences do not overlap, we formulate the detection task as a constrained  likelihood optimization problem, and design a computationally efficient dynamic program that attains its optimal solution.
	Our proposed framework is scalable, simple to implement, and robust to model uncertainties.
	We show by extensive numerical experiments that our algorithm accurately estimates the locations in dense and noisy environments, and outperforms alternative methods.
\end{abstract}

\section{Introduction}
\label{sec:introduction}
This paper  studies the classical problem of detecting signal occurrences in a one-dimensional, noisy measurement. 
This detection problem  appears in various signal processing applications, such as 
defects detection~\cite{tsai2003fast}, radar detection~\cite{levanon2005cross},   fluorescence imaging~\cite{evanko2009primer}, ultrasound imaging~\cite{tur2011innovation,bendory2016stable}, signal synchronization for communication~\cite{797329}, and GPS~\cite{8002960}. 
In particular, the main motivation of this paper arises from the task of particle picking in single-particle cryo-electron microscopy (cryo-EM):  a leading technology to constitute the three-dimensional structure of biological molecules~\cite{frank2006three,bendory2020single,singer2020computational}. The goal of particle picking is to detect the location of particle images in a noisy measurement. 
%\amichai{specifically, our signal detection problem focuses on a setup where  
	%\begin{itemize}
	%   \item The signals shapes are fixed and known.
	%  \item Signals do not coincide.
	% \item The measurement follows an additive Gaussian noise model.
	%\end{itemize}}
	%The studied 
	This problem is especially challenging   since the sought particle images might be densely packed and the signal-to-noise ratio (SNR) is  low~\cite{heimowitz2018apple,bepler2019positive,eldar2020klt};
	our model can be viewed as a one-dimensional version of this task.  
	In particular, motivated by cryo-EM, we focus on detecting fixed and non-overlapping signals, contaminated by additive Gaussian noise. 
	
	Let $y\in\R^N$ be a measurement of the form 
	\begin{equation} \label{eq:model}
		y[n] = \sum_{k=1}^{K} x[n-n_k] + \varepsilon[n],
	\end{equation}
	where $n_1,\ldots,n_K$ are the unknown locations we aim to estimate, 
	$x\in\R^L$ is the signal, and $\varepsilon[n]\sim\N(0,\sigma^2)$ is i.i.d.\ Gaussian noise.
	In Section~\ref{sec:dynamic_program} we first assume that the signal $x$, the noise level $\sigma^2$, and the number of signal occurrences $K$ are known.
	Later, in Section~\ref{sec:gap}, we extend the method to account for an unknown number of signal occurrences. In Section~\ref{sec:Experiments} we demonstrate numerically that the method is also robust to uncertainties in the signal's length. We allow the locations of the signal occurrences to be arbitrarily spread in the measurement, with a single restriction: the signal occurrences do not overlap, namely,
	\begin{equation} \label{eq:separation_condition}
		|n_i-n_j|\geq L \quad \text{for all} \quad  i\neq j;
	\end{equation}
	we refer to this restriction as the separation condition. We also define another separation condition for well-separated signals, where the signals are spaced with a minimum distance of a full signal length from each other, namely,
	\begin{equation} \label{eq:separation_condition_well_separated}
		|n_i-n_j|\geq 2L \quad \text{for all} \quad  i\neq j.
	\end{equation}

	Assuming  the noise level $\sigma^2$, the signal  $x$, and $K$ are known,  maximizing the likelihood function of~\eqref{eq:model} is equivalent to the least squares problem
	\begin{equation*}	
		\begin{split}
			\arg\min_{\hat{n}_1 ,\ldots,\hat{n}_K }& \left\|y-\sum_{k=1}^{K} x[n-\hat{n}_K]\right\|^2_2.
		\end{split}
	\end{equation*}
	Thus, it can be readily seen that  maximizing the likelihood function under the separation condition~\eqref{eq:separation_condition} is equivalent to the constrained optimization problem:
	\begin{equation}	\label{eq:likelihood}
		\begin{split}
			&\arg\max_{\hat{n}_1,\ldots,\hat{n}_K} \sum_{k=1}^K\sum_{n=0}^{N-1-L} y[n]x[n-\hat{n}_K]\\ & \;\text{subject to}\;\;   |\hat{n}_i-\hat{n}_j|\geq L \quad \text{for all} \quad i\neq j.
		\end{split}
	\end{equation}
	Solving this optimization problem accurately and efficiently is the main focus of this paper. 
	%\amichai{It is important to emphasize that if the measurement noise $\epsilon[n]$ does not follow a normal distribution, then the  proposed objective (\ref{eq:likelihood}) does not hold a maximum likelihood interpretation. Nevertheless, it is still of high interest to the cryo-EM literature, even in this scenario (Tamir, please add a reference here)}.   
	
\begin{figure*}[h]
		\centering
		\includegraphics[width=.8\linewidth]{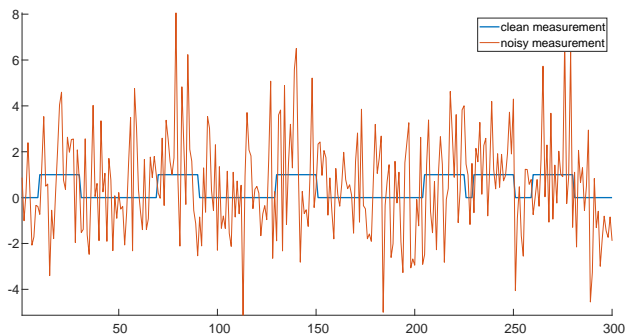}
		\caption{An example  of a clean measurement ($\sigma^2=0$) and a noisy measurement  with $\sigma^2=2$. 
		}
		\label{fig:example} 
	\end{figure*}

	Figure~\ref{fig:example} demonstrates an example of a clean measurement ($\sigma^2=0$) and a noisy measurement  with $\sigma^2=2$. 
	The clean measurement consists of six signal occurrences. 
	Note that the three signal occurrences on the left are well separated. In this regime, the detection problem is rather easy. 
	On the contrary,  the three signal occurrences on the right   are densely packed, rendering the signal detection problem challenging.  Our goal is to estimate the locations of the signal occurrences, accurately and efficiently, in both regimes.

	If the signal occurrences are well-separated (the left end of Figure~\ref{fig:example}), 
	the signal locations may be detected using the following greedy approach.
	First,  the measurement is correlated with the signal $x$ (assumed to be known) and the index corresponding to the maximum of the correlation is chosen as the first estimator $\n_1$. 
	Next,~$\n_2$ is chosen as the index corresponding to the maximum of the correlation, where the maximum is taken 
	among all entries which are separated by at least~$L$ entries from $\n_1$.
	The same strategy  is applied consecutively to estimate  $\n_3,\ldots,\n_K$.  
	Hereafter, we refer to this algorithm as the \textit{greedy algorithm}.
	This algorithm is highly efficient, as the correlations may be executed with only a few FFTs~\cite{4428580}. The greedy approach is a very popular scheme in many real-world applications~\cite{doi:10.1080/13102818.2017.1389303,fukunishi2018improvements}.
	However, this simple approach fails in cases where the signals are close, as demonstrated in the right end of Figure~\ref{fig:example}. 
	
	The main contribution of this paper is an exact and efficient algorithm to maximize the likelihood function~\eqref{eq:likelihood}. 
	In Section~\ref{sec:dynamic_program} we describe how this maximum is attained by utilizing dynamic programming. Based on the principle of gap statistics, Section~\ref{sec:gap} extends the scope of our problem and studies the case where the number of signal occurrences $K$ is unknown.
	In Section~\ref{sec:Experiments} we conduct comprehensive  numerical experiments to study the performance of the proposed dynamic program, its robustness, and compare it to the greedy algorithm.
	In Section~\ref{sec:convex}, we also compare the dynamic program with a convex program that was developed in the context of super-resolution~\cite{candes2014towards,bendory2016robust,morgenshtern2016super,bendory2017robust}.
	Finally, we conduct a few experiments on one-dimensional stripes of cryo-EM data (the original data is two-dimensional), indicating that the dynamic program can estimate the locations of densely packed particle images, while the greedy algorithm fails. 
	We conclude the paper in Section~\ref{sec:conclusions} by discussing the challenges of extending this framework to two-dimensional data, such as cryo-EM data sets.

	\section{Dynamic programming for signal detection} \label{sec:dynamic_program}
	Dynamic programming is a method for breaking down a problem into simpler sub-problems and solving them recursively~\cite{bellman1954theory}.
	In particular, our proposed dynamic program solution is based on the following procedure.
	Let $g[n,j]$ be the maximum of~\eqref{eq:likelihood}, for~$j$ signal occurrences, over indices $1,\dots,n$. By definition, $g[N,K]$ is the sought solution of~\eqref{eq:likelihood}. Our proposed dynamic program rule is given by:
	\begin{align}
		\label{DP rule}
		g[n+1,&j]=\max\left\{g[n,j], g[n-d,j-1]+f[n+1]\right\},
	\end{align}
	where 
	\begin{equation} \label{eq:f}
		f[m]=\sum_{i=m-d}^{m+d}y[i]x[i-(m-d)],
	\end{equation}
	is the correlation between the measurement $y$ at the interval $[m-d,m+d]$  and the signal~$x$, while $d=\lfloor L/2 \rfloor$ is half the length of the signal. 
	In words, the maximal objective for locating $j$ signals over indices  $1,\dots,n+1,$ is the maximum between the following two options:
	\begin{enumerate}
		\item the best we can achieve for locating $j$ signals over indices  $1,\dots,n$ (namely, the solution of the previous step);
		\item the best we can achieve under the constraint that a signal is located at  location $n+1$.
	\end{enumerate}
	
	The dynamic program rule introduces a simple \textit{bottom up} routine for finding the maximum of the  objective $g[N,K]$. That is, we define a matrix $g$ of dimensions $N\times K$, where each row corresponds to the indices of the measurement and each column is the number of signal occurrences. Then, we iterate over $j=1,\dots,K$ and $i=1,\dots,N$, and fill the entries $g[i,j]$ according to~\eqref{DP rule}. Notice that for every $g[i,j]$, we also store the corresponding estimated signal locations $\hat{n}_1,\dots,\hat{n}_j$.
	Finally, we return $g[N,K]$, and the corresponding signal locations, as desired.
	Algorithm~\ref{alg:Ours} summarizes our proposed scheme. Notice that a signal cannot be located near the staring and end indices, namely, at $i<L$ and $i>N-L$. This means that $g[i,j]=0$ for all $i<L$ and $g[i,j]=g[i-1,j]$ for all $i>N-L$. We exclude these cases from the description of Algorithm~\ref{alg:Ours} for  simplicity of presentation. 
	
	\begin{algorithm}[h]
		
		\begin{algorithmic}[1]
			\renewcommand{\algorithmicrequire}{\textbf{Input:}}
			\renewcommand{\algorithmicensure}{\textbf{Output:}}
			\REQUIRE  $y\in \mathbb{R}^N$, $x\in\mathbb{R}^{L}$, and the number of signal occurrences $K$ 
			\FOR {$k=1$ to $K$}
			\FOR {$i=1$ to N}
			\STATE 
			compute $g[i,k]$ according to~\eqref{DP rule}
			\ENDFOR 
			\ENDFOR
			\RETURN $g[N,K]$ and estimates of the locations of the signal occurrences $\hat n_1,...\hat{n}_K$
		\end{algorithmic}
		\caption{Signal detection using dynamic programming}
		\label{alg:Ours}
	\end{algorithm}

	\begin{algorithm}
		\begin{algorithmic}[1]
			\renewcommand{\algorithmicrequire}{\textbf{Input:}}
			\REQUIRE  $y\in \mathbb{R}^N$, $x\in\mathbb{R}^{L}$, and the number of signal occurrences  $K$
			%		\ENSURE  ${{\hat n_1,...,\hat n_{K}}}$
			\STATE compute $f$ according to~\eqref{eq:f}
			\FOR {$i=1$ to $K$}
			
			\STATE 
			$\n_i = \, \arg\max \, f$ \quad  subject to \quad  $|\n_i - \n_j|\geq L$ for all $j<i$
			\STATE 
			$\gamma_K[i] = \, \max f \,$ \quad subject to \quad  $|\n_i - \n_j|\geq L$ for all $j<i$
			
			\ENDFOR
			\STATE $\gamma_K = \sum_{i=1}^K \gamma_K[i]$
			\RETURN  Estimates of the locations of the signal occurrences $\hat n_1,...\hat{n}_K$, and $\gamma_K$
		\end{algorithmic}
		\caption{Signal detection using the greedy approach}
		\label{alg:Greedy}
	\end{algorithm}

	The computational complexity of our algorithm is \textit{O}($N\cdot \max\{K,\log{N}\}$), as follows. Computing the cross correlation between $y$ and $x$ costs \textit{O}($N \log N$) operations using the  FFT algorithm. 
	Given the cross correlation values, every iteration of Algorithm~\ref{alg:Ours} is of~\textit{O}($1$).
	Overall, we have \textit{O}$(NK)$ iterations, and thus the computational complexity of the entire proposed scheme is  \textit{O}($N\cdot \max\{K,\log{N}\}$). 
	
	In Sections~\ref{sec:Experiments} and \ref{sec:Cryo Experiment}, we compare Algorithm~\ref{alg:Ours} against the greedy algorithm described in Section~\ref{sec:introduction}.
	This algorithm chooses the peaks of the cross correlation between the signal and the measurement, while forcing a separation. Thus, its computational complexity  is  \textit{O}($N\log{N}+K$). For small $K$, the complexities of both algorithms match. 
	Algorithm~\ref{alg:Greedy} summarizes this method. 
	
	We mention in passing that our problem shares some similarities with the change point detection problem---a well-studied problem in statistics. 
	One popular solution to change point detection is based on dynamic programming~\cite{auger1989algorithms,rigaill2015pruned}.
	Yet, this algorithm is significantly different from the dynamic program in Algorithm~\ref{alg:Ours}.
	
	\section{Estimating the number of signal occurrences using the gap statistics principle} \label{sec:gap}
	
	In many real-world applications, the number of signal occurrences $K$ is unknown. This problem is of special interest, as both the greedy algorithm and our proposed scheme assume knowledge of $K$. The classical approach for finding $K$ is based on finding a ``knee''  behavior (also referred to as the \textit{elbow  method}). 
	This heuristic suggests solving \eqref{eq:likelihood} for different values of $K$, and returning the value that introduces the steepest decrease in the objective value. This approach is perhaps the most popular framework in many related  applications, such as clustering {~\cite{tibshirani2001estimating}}, regularization ~\cite{doi:10.1137/1034115} and others. It was extensively studied and improved over the years, see for example~\cite{bholowalia2014ebk,epub11920,kou2014estimating}. 
	
	In our work, we suggest using the principle of gap statistics: a statistically-driven modification of the knee approach, which was first introduced in~\cite{tibshirani2001estimating} in the context of estimating the optimal number of clusters in a data set. 
	In their work, Tibshirani et al. \cite{tibshirani2001estimating} showed that in standard clustering, the error measure monotonically decreases as the number of clusters increases, however, from some value of $K$ onward, the decrease flattens markedly. This $K$ is usually referred to as the ``knee'' of the plot, and is believed to indicate the optimal number of clusters  in the data set.
	
	The gap statistic provides a statistical procedure to formulate the detection of this knee. The key idea of this approach is to standardize the curve of the objective value  by comparing it with its expectation under an appropriate null reference. 
	Formally,  Tibshirani et al. defined the gap statistic as
	\begin{equation*}
		\gap_N(K)=\mathbb{E}_N^*w(K)-w(K),
	\end{equation*}
	where $w(K)$ is the objective value over $K$ clusters, and $\mathbb{E}_N^*$ denotes the expectation under a sample of size $N$ from a ``null'' reference distribution. By ``null'' we mean a clustering performed on noise.
	The estimate of the number of classes, denoted by $\hat{K}$, is the value that maximizes $\gap_N(K)$. Intuitively, $\hat{K}$ implies the ``strongest'' evidence against the null. The Gap statistic was extensively studied and applied to many applications~\cite{rathod2017design,wang2016extended,zheng2009estimating,arima2008modified,painsky2012exclusive,painsky2014optimal}. 
	
	In this work, we take a similar approach, and suggest estimating the number of signal occurrences based on the gap statistics principle. First, we observe that~\eqref{eq:likelihood} is monotonically 
	increasing in~$K$, where the steepest decrease is expected at the vicinity of the true value of~$K$. 
	For a given measurement $y$ and a range of values of~$K$, we apply  Algorithm~\ref{alg:Ours} and find the maximal objective value $g(N,K)$.
	We note that no additional computations are required at this stage, since the dynamic program already computes $g(N,j)$ for $j=1,\ldots,K.$
	To apply the gap statistics procedure, we also need to 
	evaluate $\mathbb{E}_N^*g(N,K)$ for every $K$. That is, the expected objective under the null. Here, we define the null as the case where no signal is embedded in the measurement. Therefore, to approximate $\mathbb{E}_N^*g(N,K)$ we simply permute the vector~$y$  $P$ times, drawn i.i.d.\ from a uniform distribution over all possible permutations, and apply Algorithm \ref{alg:Ours} on the permuted measurements, $\tilde{y}_1,\ldots,\tilde{y}_P$. Notice that by permuting the indices of $y$, we break the embedded signals (if such exist) and attain a vector with $K=0$ (henceforth, the null). Letting $g_i(N,K)$ be the value of~\eqref{DP rule} for the permuted measurement   $\tilde{y}_i$, we have $\mathbb{E}_N^*g(N,K)\approx \frac{1}{P}\sum_{i=1}^Pg_i(N,K)$ for large enough $P$. 
	Therefore, we  approximate the statistical gap by 
	\begin{equation}
		\label{gap_statistic}
		\gap_N(K)\approx\frac{1}{P}\sum_{i=1}^Pg_i(N,K)-g(N,K),
	\end{equation}
	for every $K$, and return the value of $K$ which maximizes the gap (similarly to~\cite{hastie2001estimating}). We summarize our approach in Algorithm \ref{alg:Gap Statistic K find}. 
	Algorithm~\ref{alg:Gap Statistic K find greedy} shows the analog of the greedy algorithm, Algorithm~\ref{alg:Greedy}, in the case where the number of signal occurrences is unknown, and estimated using the  gap statistic method.

	\begin{algorithm}[h]
		
		\begin{algorithmic}[1]
			\renewcommand{\algorithmicrequire}{\textbf{Input:}}
			\renewcommand{\algorithmicensure}{\textbf{Output:}}		
			
			\REQUIRE  $y\in \mathbb{R}^N$, $x\in\mathbb{R}^{L}$, $K_{\max}$
			\FOR {$K=1$ to $K_{\max}$}	
			\STATE Compute $g(N,K)$ with respect to $y$ using Algorithm~\ref{alg:Ours}
			
			\FOR {$i=1$ to $P$}
			\STATE Compute $g_i(N,K)$ with respect to $y_i$ (a permutation of $y$) using Algorithm~\ref{alg:Ours}		\ENDFOR 
			\STATE Compute  $\gap_N(K)=\frac{1}{P}\sum_{i=1}^Pg_i(N,k)-g(N,k)$
			\ENDFOR
			\STATE Compute $\hat{K}=\arg\max\, \gap_N(K)$
			\RETURN $\hat{K}$ (an estimate of the number of signal occurrence in the measurement), $g(N,\hat{K})$, and estimates of the locations of the signal occurrences $\hat{n}_1,\ldots,\hat{n}_{\hat{K}}$
		\end{algorithmic}
		\caption{Signal detection using dynamic programming with an unknown number of signal occurrences }
		\label{alg:Gap Statistic K find}
	\end{algorithm} 
	
	\begin{algorithm}[h]
		
		\begin{algorithmic}[1]
			\renewcommand{\algorithmicrequire}{\textbf{Input:}}
			\renewcommand{\algorithmicensure}{\textbf{Output:}}		
			
			\REQUIRE  $y\in \mathbb{R}^N$, $x\in\mathbb{R}^{L}$, $K_{\max}$
			\FOR {$K=1$ to $K_{\max}$}			
			\STATE Evaluate $\gamma_K$ with respect to $y$ using Algorithm~\ref{alg:Greedy}
			
			\FOR {$i=1$ to $P$}
			\STATE Compute $\tilde \gamma_{K,i}$ with respect to $\tilde{y}_i$ (a permutation of $y$) using Algorithm~\ref{alg:Greedy}	
			\ENDFOR 
			\STATE Compute  $\gap_N(K)=\frac{1}{P}\sum_{i=1}^P\tilde \gamma_{K,i}-\gamma_K$
			\ENDFOR
			\RETURN $\hat{K}$ (an estimate of the number of signal occurrence in the measurement),  and estimates of the locations of the signal occurrences $\hat{n}_1,\ldots,\hat{n}_{\hat{K}}$
		\end{algorithmic}
		\caption{Signal detection using the greedy approach with an unknown number of signal occurrences}
		\label{alg:Gap Statistic K find greedy}
	\end{algorithm}

	\section{Numerical experiments}
	\label{sec:Experiments} 
	
	In this section we compare the proposed dynamic program with alternative methods. We  use the $F_1$-score to evaluate the performance of the studied methods~\cite{huang2015maximum,fujino2008multi,ghaddar2018robust}. It is formally defined as 
	\begin{equation}
		\label{$F_1$}
		F_1=2\times \frac{\text{precision}\times \text{TPR}}{\text{precision}+\text{TPR}},
	\end{equation}
	where \textit{precision} is the ratio of the true positives (correct detections) over all detections, while TPR is the true positive rate; the ratio of true positives over all signal occurrences. In addition, we also report the \textit{recall}; the ratio of true positives over all positives, for completion. In practice, we cannot expect an exact detection of a signal location. Therefore, we follow \cite{schwartzman2011multiple} and declare  a true detection if  $|\hat{n}_k-n_k|<L/2$. That is, we say that a signal is correctly detected if its estimated location  is within $L/2$ entries from the  true location. This convention is quite popular in relevant signal detection literature \cite{schwartzman2011multiple,cheng2017multiple}. Further, it is well-motivated by our cryo-EM application. Specifically, in cryo-EM particle picking, the displacements are not a big issue since the images are later aligned  as part of the refinement algorithm \cite{bendory2020single}.

	We begin with synthetic experiments. We generate a measurement $y$ as follows.
	First, we fix the measurement length $N$, the number of signal occurrences $K$,  and signal length $L$. Then, we place the first signal at a random location. Next, we draw a new location; if the new location is eligible, then we place it; if not, we draw a new location. We repeat this process until we place all~$K$ signals. 
	By eligible location, we mean that the left-most point of the new signal is separated by at least~$L$ entries from the left-most point of all previous signals for an \textit{arbitrary-spaced measurement}~\eqref{eq:separation_condition} (so the signal occurrences do not overlap) and $2L$ for a \textit{well-separated measurement}~\eqref{eq:separation_condition_well_separated}. Finally, we add independent and identically distributed white Gaussian noise with zero mean and variance $\sigma^2$. The code to reproduce all experiments is publicly available at \url{https://github.com/MordechaiRoth1/Signal-detection-with-dynamic-programming}.
	
	\subsection{Performance for a known number of signal occurrences}
	\label{subsec:all known}
\begin{figure*}
		\begin{subfigure}[h]{1\columnwidth}
			\centering
			\includegraphics[width=0.6\columnwidth,bb= 20 10 1200 600,clip]{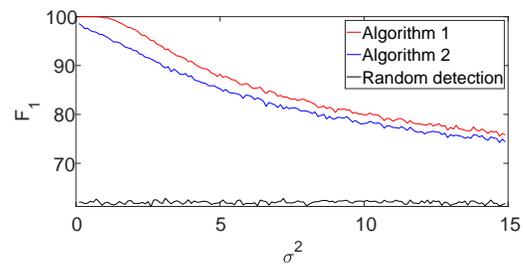}
			\caption{\label{fig:dense all known} Arbitrary spaced setup}
		\end{subfigure}
		\hfill
		\begin{subfigure}[h]{1\columnwidth}
			\centering
			\includegraphics[width=0.6\columnwidth,bb= 20 10 1200 600,clip]{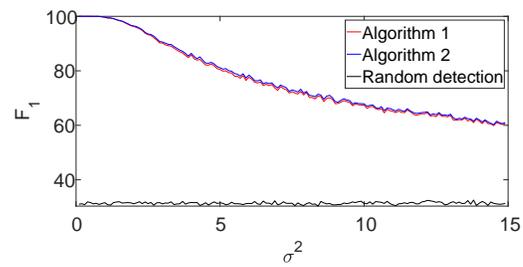}
			\caption{\label{fig:well separated all known} Well separated setup}
		\end{subfigure}
		\caption{\label{fig:all known} $F_1$-score for Algorithm~\ref{alg:Ours} and Algorithm~\ref{alg:Greedy} for the arbitrary spaced and well separated setups, assuming the signal's shape and the  number of  occurrences are known. 
		}
	\end{figure*}

	First, we compare the  the performance of our dynamic programming scheme (Algorithm~\ref{alg:Ours}) with the greedy approach (Algorithm~\ref{alg:Greedy}) in the ideal case, where the signal's shape and the number of signal occurrences are known. We use a rectangular signal of length $L=30$, where all of its entries are equal to one. We place them in a measurement of length $N=300$, as described above. For the well-separated setup, we set $K=3$ signal occurrences that satisfy the separation condition~\eqref{eq:separation_condition_well_separated} and for the arbitrary spaced case, we use $K=6$ signal occurrences, that only satisfy the $L$-separation condition~\eqref{eq:separation_condition}.  For each noise level $\sigma^2$, we conduct $3000$ trials, each with a fresh measurement. As a baseline, we further compute the $F_1$-score of a random detection process, where $K$ locations which satisfy~\eqref{eq:separation_condition}  are chosen at random. 
	Figure~\ref{fig:all known} presents the results.
	First, it is evident that
	the performance of the algorithms are comparable for the well separated case in Figure~\ref{fig:well separated all known}. However, we observe that the dynamic program outperforms the greedy algorithm  in cases where the signals are dense as in Figure~\ref{fig:dense all known}. The complementary recall charts are quite similar to the $F_1$ scores and are provided in the Appendix. 
	
	\subsection{Performance for an unknown number of signal occurrences}
	\label{subsec:unknownK}
	
\begin{figure*}
		\begin{subfigure}[h]{1\columnwidth}
			\centering
			\includegraphics[width=0.6\columnwidth,bb= 20 10 1200 600,clip]{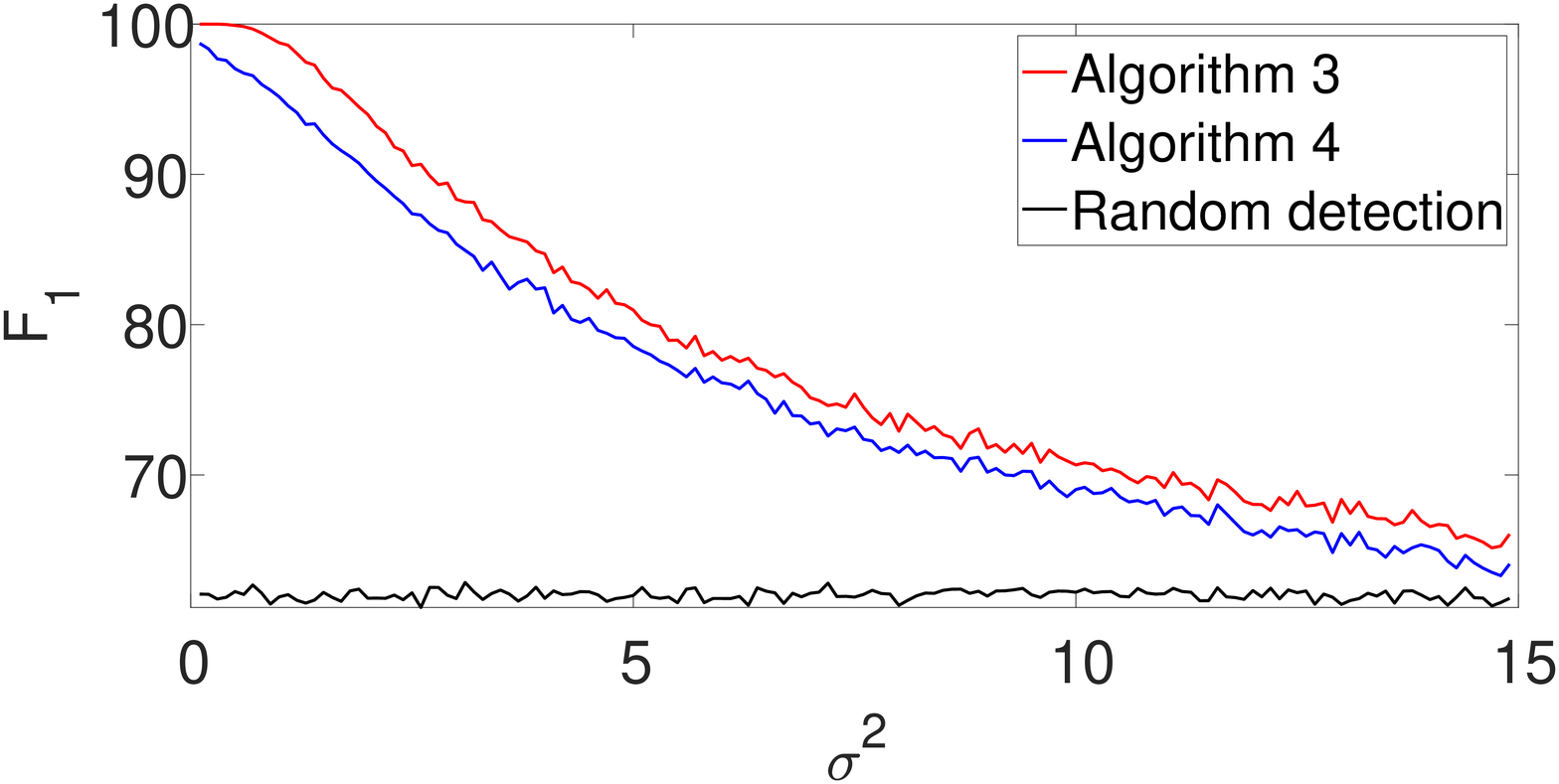}
			\caption{\label{fig:arbitrary spaced dense all known} Arbitrary spaced setup}
		\end{subfigure}
		\hfill
		\begin{subfigure}[h]{1\columnwidth}
			\centering
			\includegraphics[width=0.6\columnwidth,bb= 20 10 1200 600,clip]{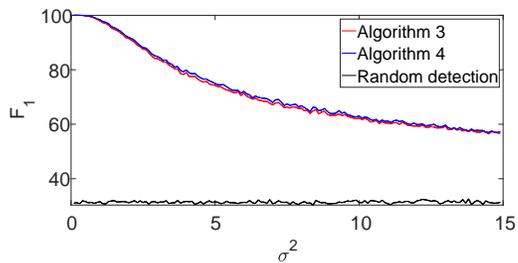}
			\caption{\label{fig:well separated  unknownK} well separated setup}
		\end{subfigure}
		\caption{\label{fig:L known} $F_1$-score for Algorithm \ref{alg:Gap Statistic K find} and for Algorithm \ref{alg:Gap Statistic K find greedy} 
			for the arbitrary spaced and well separated setups, assuming the signal's shape is known but the   number of  signal occurrences $K$ is unknown.}
	\end{figure*}

	Next, we repeat the previous experiment while dropping the known $K$ assumption (yet, the signal's shape is still assumed to be known). 
	In this case, we apply the gap statistic principle to evaluate the number of signal occurrences, while estimating their locations as described in Algorithms~\ref{alg:Gap Statistic K find} and \ref{alg:Gap Statistic K find greedy}.  The  results are presented in Figure~\ref{fig:L known}. As in the previous example, the performance of both algorithms is comparable for the well separated case (Figure~\ref{fig:well separated  unknownK}), while the dynamic program is clearly superior in the arbitrarily spaced (henceforth, dense) setup (Figure~\ref{fig:arbitrary spaced dense all known}). 
	As expected, the performance of the algorithms deteriorates compared to Figure~\ref{fig:all known}. Once again, we report the recall in the Appendix as it demonstrates quite a similar behavior. 
	
	%It is visible in figure \ref{fig:L known} that Algorithm \ref{alg:Gap Statistic K find} maintains superiority over \ref{alg:Gap Statistic K find greedy} for arbitrarily spaced setups while maintaining similar results for well separated setups.
	
\begin{figure*}
		\begin{subfigure}[h]{1\columnwidth}
			\centering
			\includegraphics[width=0.6\columnwidth,bb= 20 10 1200 600,clip]{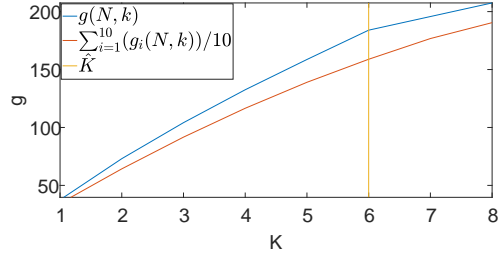}
			\caption{\label{fig:GapOurs}Gap found by Algorithm \ref{alg:Gap Statistic K find} }
		\end{subfigure}
		\hfill
		\begin{subfigure}[h]{1\columnwidth}
			\centering
			\includegraphics[width=0.6\columnwidth,bb= 20 10 1200 600,clip]{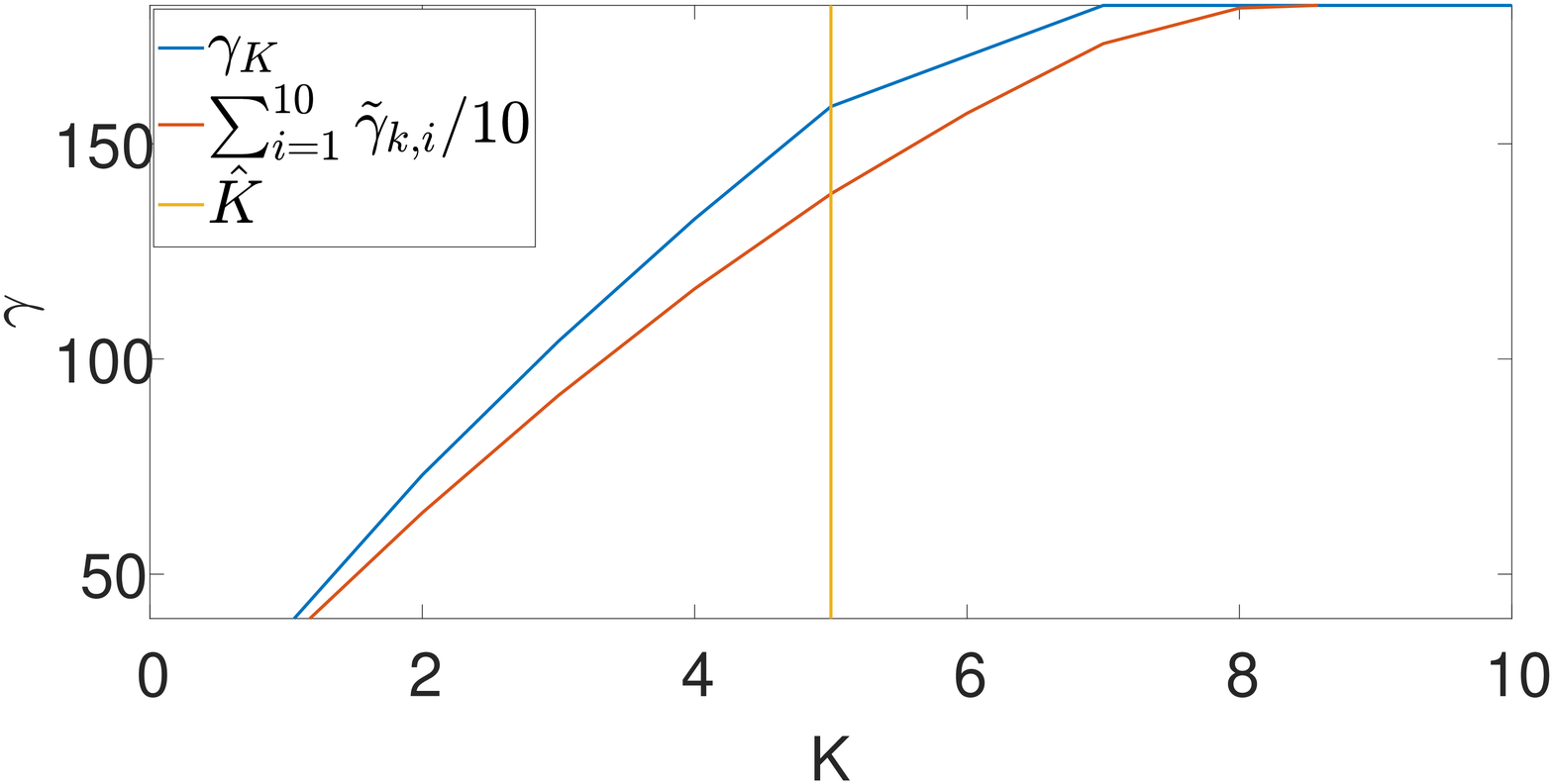}
			\caption{\label{fig:fig:GapGreedy} Gap found by Algorithm \ref{alg:Gap Statistic K find greedy}}
		\end{subfigure}
		\caption{\label{fig:Gap Example} An illustration of the gap statistic principle.  The blue curve is the measured objective, while the red curve corresponds to the approximated null. The yellow line is the maximal gap between the two and, henceforth the estimated $K$.}
	\end{figure*}

	Further, we illustrate our proposed gap statistic scheme in Figure \ref{fig:Gap Example}. Here, we set the (unknown) number of occurrences as $K=6$. The blue curve corresponds to the objective value, while the red curve is the approximated null (see \eqref{gap_statistic}). The yellow line corresponds to the maximum gap between the two curves, which is the estimated  $K$. As we can see, the gap statistic demonstrates a relatively accurate estimate of the true number of signal occurrences, in both algorithms. 
	
	\subsection{Performance with unknown signal length}
	\label{sec: unknown All}
\begin{figure*}
		\begin{subfigure}[h]{1\columnwidth}
			\centering
			\includegraphics[width=0.7\columnwidth,bb= 20 10 1200 600,clip]{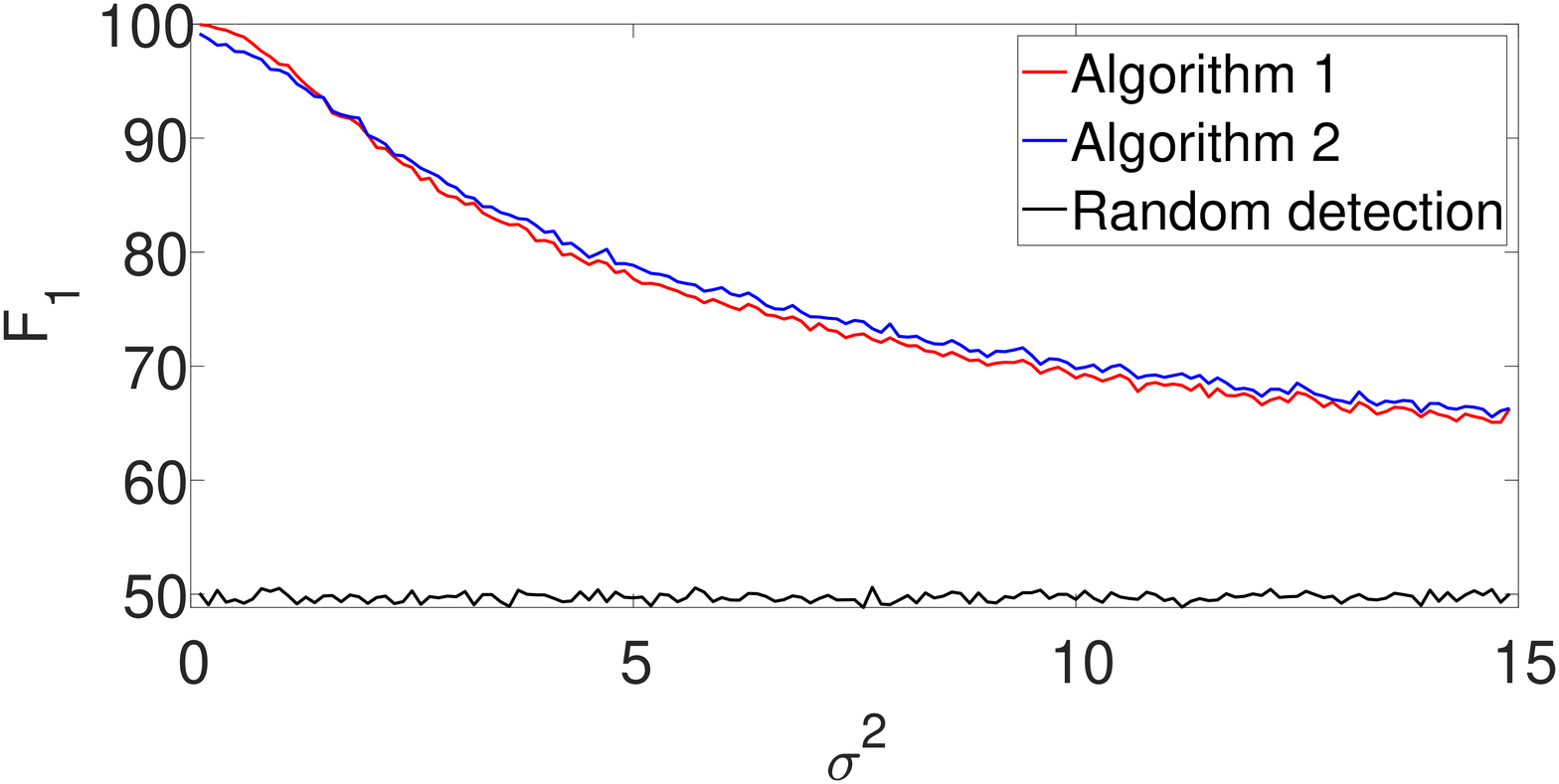}
			\caption{\label{fig:denseKnownK08} $\hat{L}/L=0.8$, known $K$}
		\end{subfigure}
		\hfill
		\begin{subfigure}[h]{1\columnwidth}
			\centering
			\includegraphics[width=0.7\columnwidth,bb= 20 10 1200 600,clip]{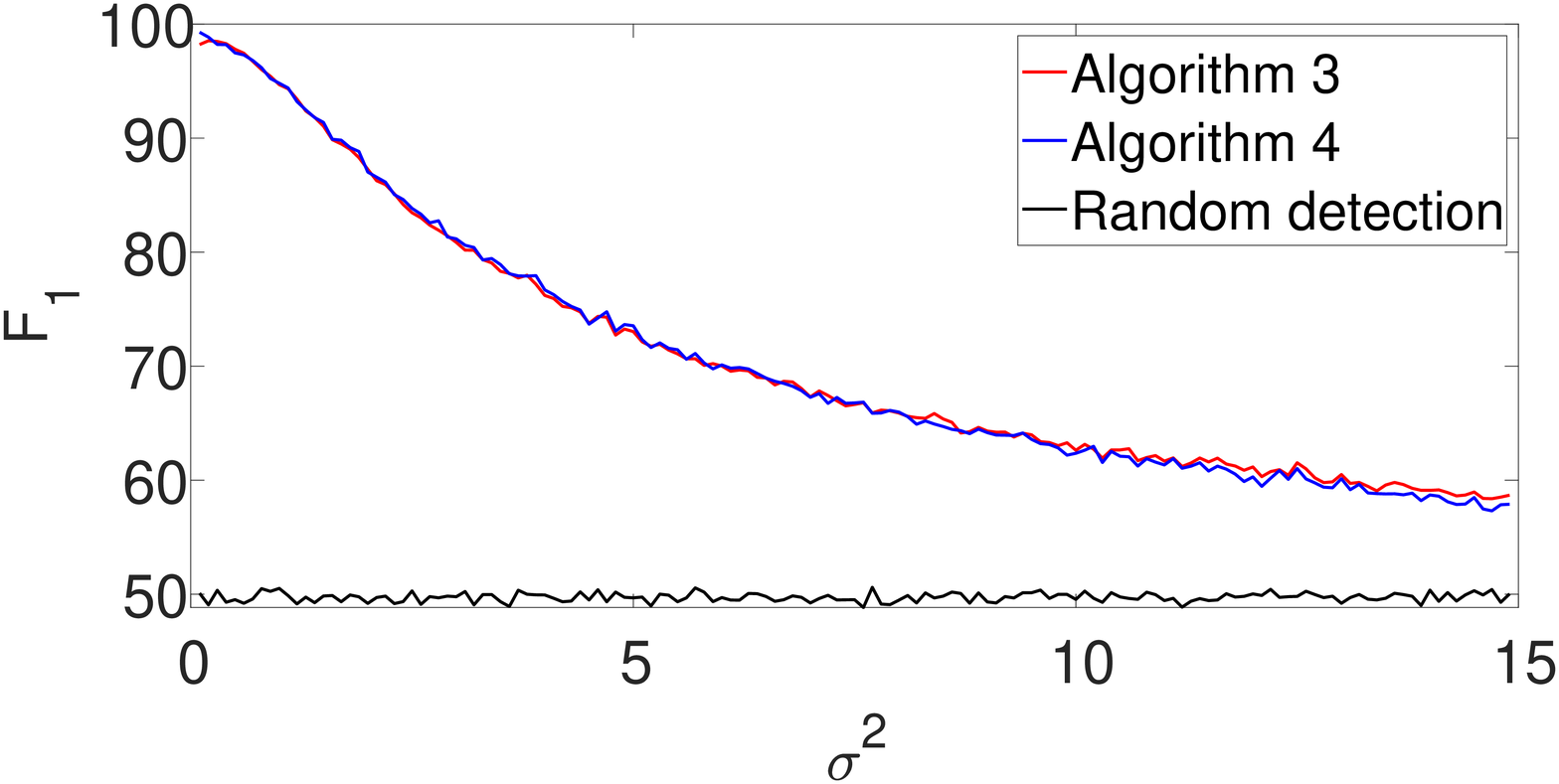}
			\caption{\label{fig:denseUnknownK08} $\hat{L}/L=0.8$, unknown $K$}
		\end{subfigure}
		\hfill

		\begin{subfigure}[h]{1\columnwidth}
			\centering
			\includegraphics[width=0.7\columnwidth,bb= 20 10 1200 600,clip]{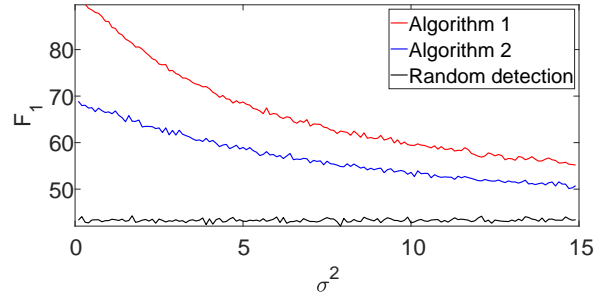}
			\caption{\label{fig:denseKnownK13} $\hat{L}/L=1.3$, known $K$}
		\end{subfigure}
		\hfill
		\begin{subfigure}[ht]{1\columnwidth}
			\centering
			\includegraphics[width=0.7\columnwidth,bb= 20 10 1200 600,clip]{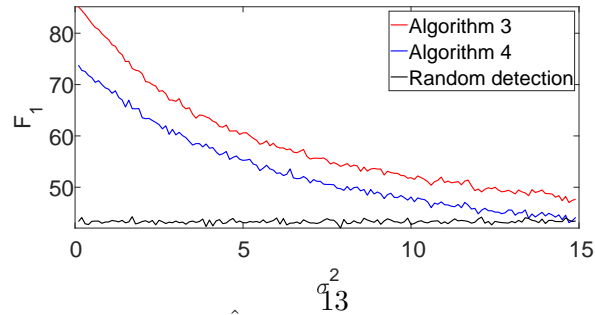}
			\caption{\label{fig:denseUnknownK13} $\hat{L}/L=1.3$, unknown $K$}
		\end{subfigure}

		\caption{\label{fig:RobustLengthDense} $F_1$-score for the arbitrary spaced model  where the signal length is unknown. Here, $\hat{L}$ denotes the assumed length of the signal. 
		}
		
	\end{figure*}

\begin{figure*}
		\begin{subfigure}[h]{1\columnwidth}
			\centering
			\includegraphics[width=0.7\columnwidth,bb= 20 10 1200 600,clip]{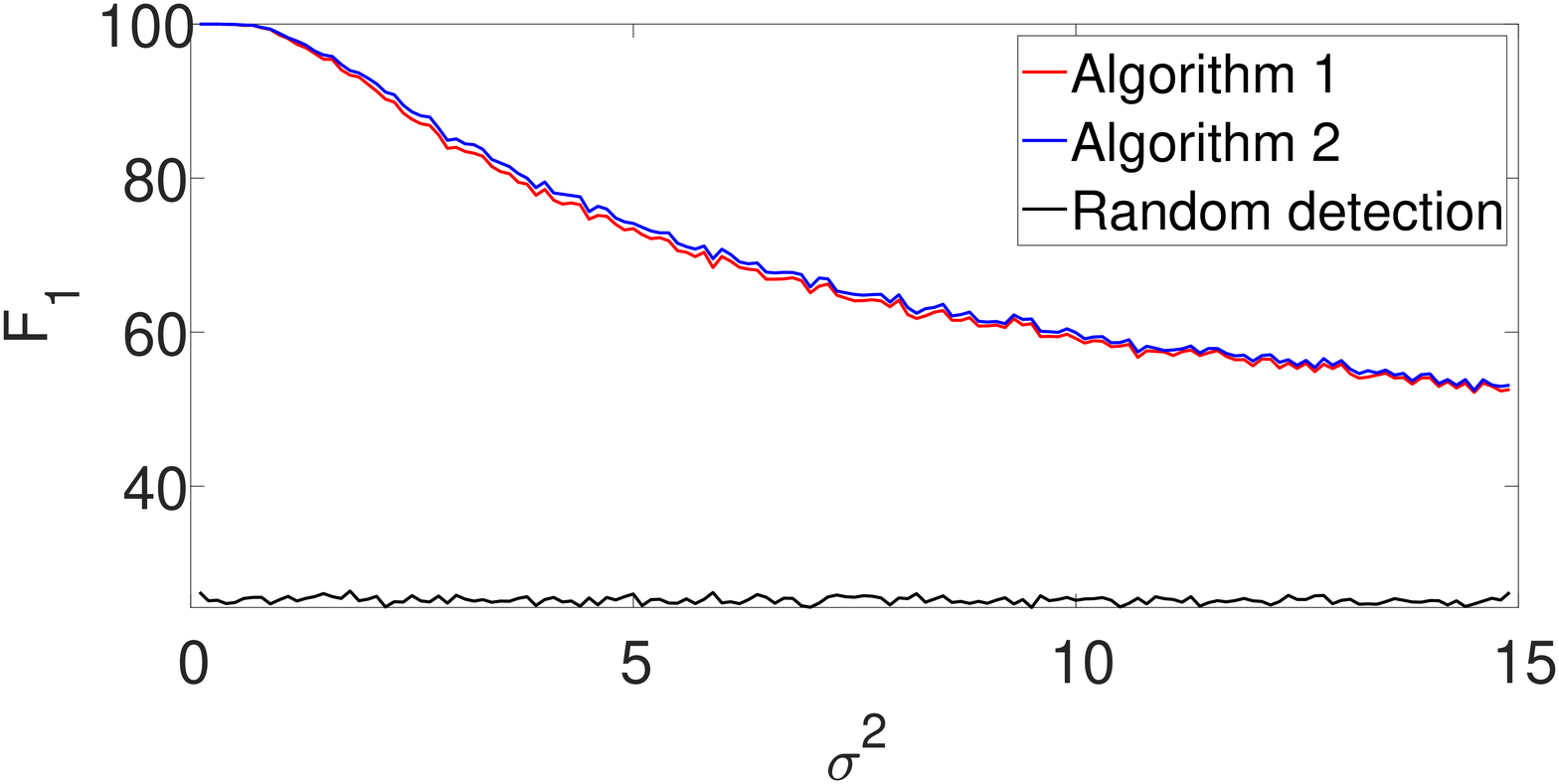}
			\caption{\label{fig:sparseKnownK08} $\hat{L}/L=0.8$, known $K$}
		\end{subfigure}
		\hfill
		\begin{subfigure}[h]{1\columnwidth}
			\centering
			\includegraphics[width=0.7\columnwidth,bb= 20 10 1200 600,clip]{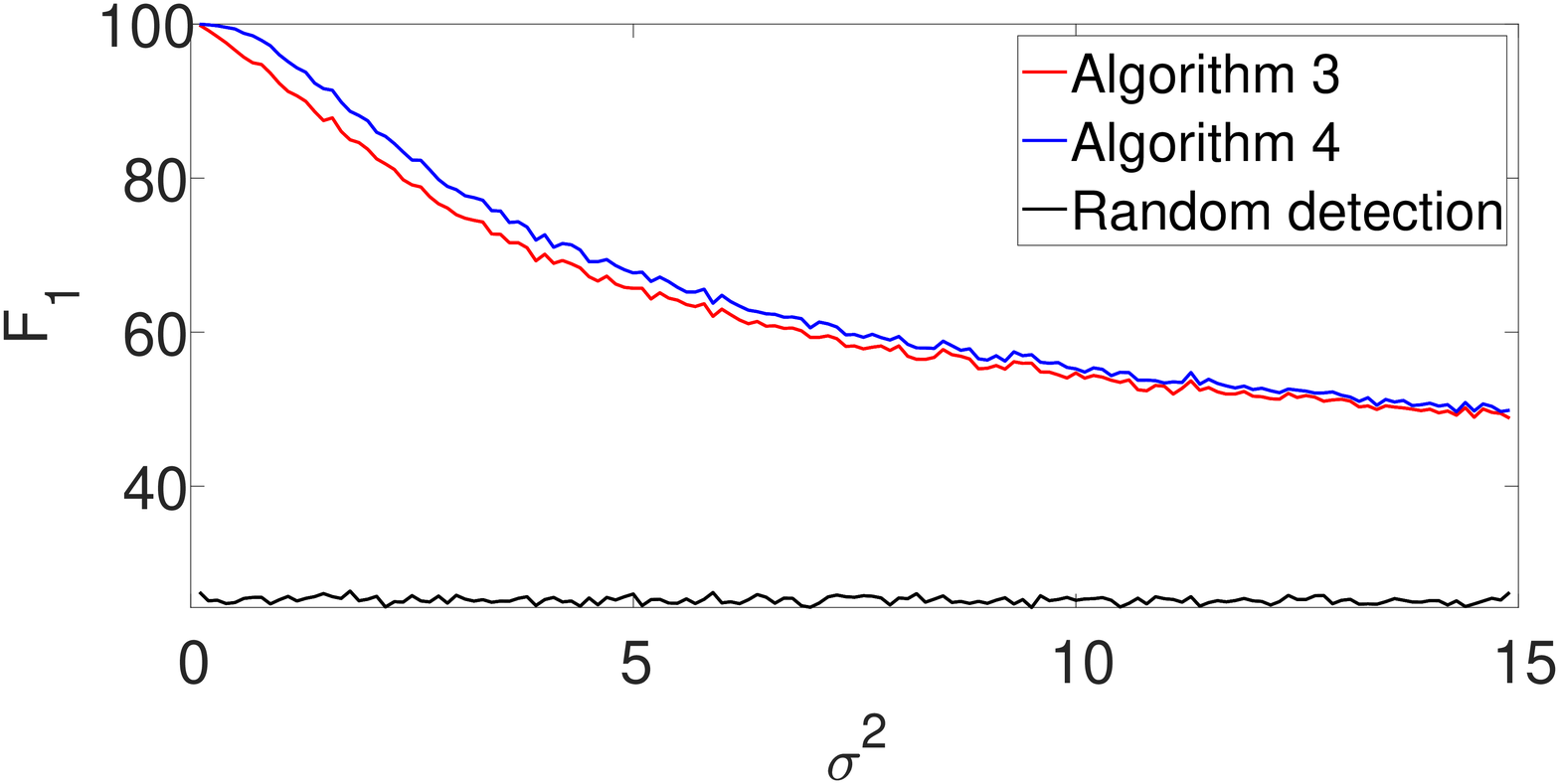}
			\caption{\label{fig:sparseUnknownK08} $\hat{L}/L=0.8$, unknown $K$}
		\end{subfigure}
		\hfill
		
		\begin{subfigure}[h]{1\columnwidth}
			\centering
			\includegraphics[width=0.7\columnwidth,bb= 20 10 1200 600,clip]{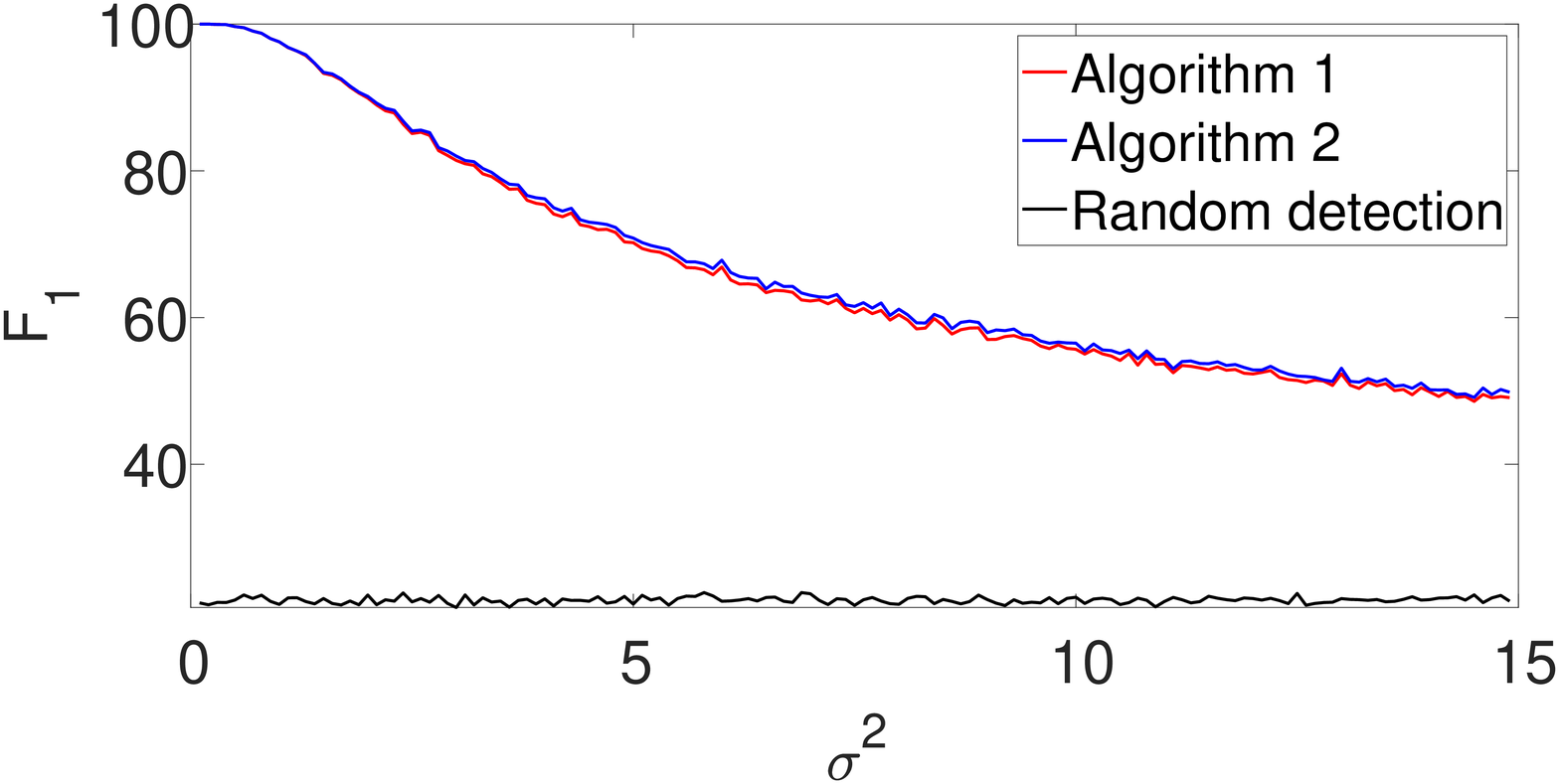}
			\caption{\label{fig:sparseKnownK13} $\hat{L}/L=1.3$, known $K$}
		\end{subfigure}
		\hfill
		\begin{subfigure}[ht]{1\columnwidth}
			\centering
			\includegraphics[width=0.7\columnwidth,bb= 20 10 1200 600,clip]{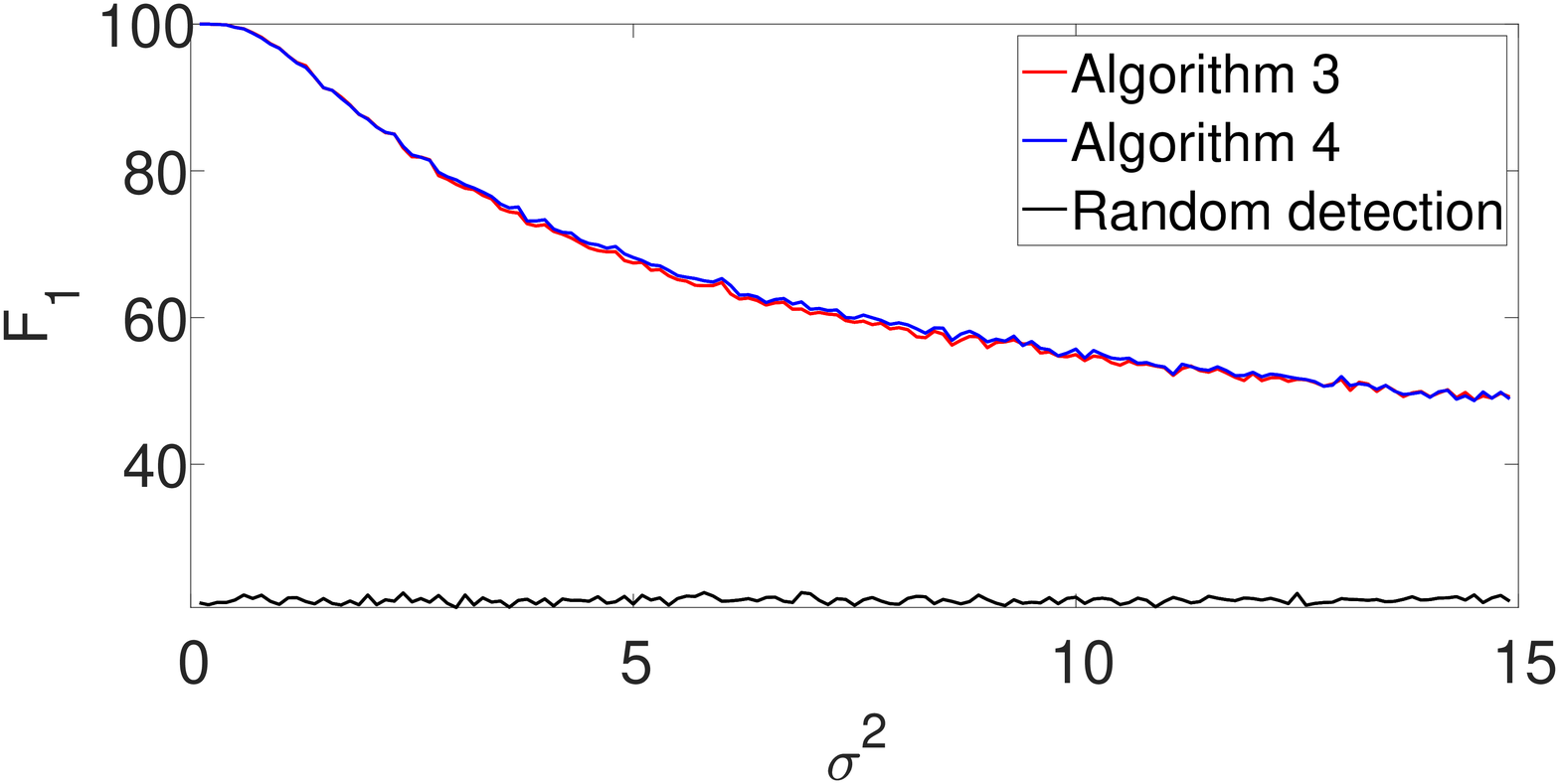}
			\caption{\label{fig:sparseUnknownK13} $\hat{L}/L=1.3$, unknown $K$}
		\end{subfigure}

		\caption{\label{fig:RobustLengthSparse} 
			$F_1$-score for the well separated model  where the signal length is unknown. 
		}
	\end{figure*}

	Further, we study the robustness of our proposed scheme, as we focus on the case where the length of the signal~$L$ is not precisely known. Let  $\hat{L}$ denote the approximated signal length. We examine two cases: $\hat{L}/L=0.8$ (the true signal's length is greater than its approximation) and $\hat{L}/L=1.3$. 
	We study the performance of our suggested framework in cases where $K$ is either known or unknown. The $F_1$ results are presented in Figure~\ref{fig:RobustLengthDense}  and Figure~\ref{fig:RobustLengthSparse} for the arbitrary spaced and well separated cases, respectively. The complementary recall charts are again reported in the Appendix.  
	
	In the arbitrary spaced setup, we observe a similar behavior for $\hat{L}/L = 0.8$ (Figures~\ref{fig:denseKnownK08} and \ref{fig:denseUnknownK08}), while our proposed method outperforms the greedy algorithm for $\hat{L}/L = 1.3$ (Figures~\ref{fig:denseKnownK13} and \ref{fig:denseUnknownK13}).  
	The reason for this phenomenon can be explained as follows.
	When $\hat{L}/L = 1.3$, the true signal is shorter than assumed. 
	Thus, the greedy algorithm declares close signals as a single realization. For $\hat{L}/L = 0.8$, both algorithms perform quite the same, as our proposed algorithm does not impose a strong enough separation constraint. 
	
	Figure~\ref{fig:RobustLengthSparse} shows the $F_1$-score for the well separated setup. Here, the performance of the greedy algorithm is comparable to the dynamic program, in all the examined setups. This is behavior is not surprising. In the well separated regime, the separation constraint is less effective, and both algorithms perform quite similarly, regardless to the accuracy of $\hat{L}$. 
	
	\subsection{Performance as a function of the measurement length}
	\label{subsec:length_increase}
	
\begin{figure*}
		\begin{subfigure}[h]{\columnwidth}
			\centering
			\includegraphics[width=0.7\columnwidth,bb= 20 10 1200 600,clip]{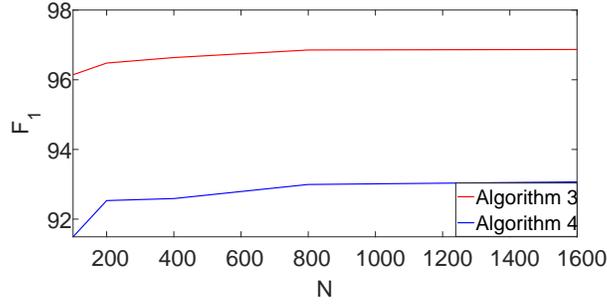}
			\caption{\label{fig:LengthIncreaseF} $F_1$-score as a function of $N$}
		\end{subfigure}
		\begin{subfigure}[h]{\columnwidth}
			\centering
			\includegraphics[width=0.7\columnwidth,bb= 20 10 1200 600,clip]{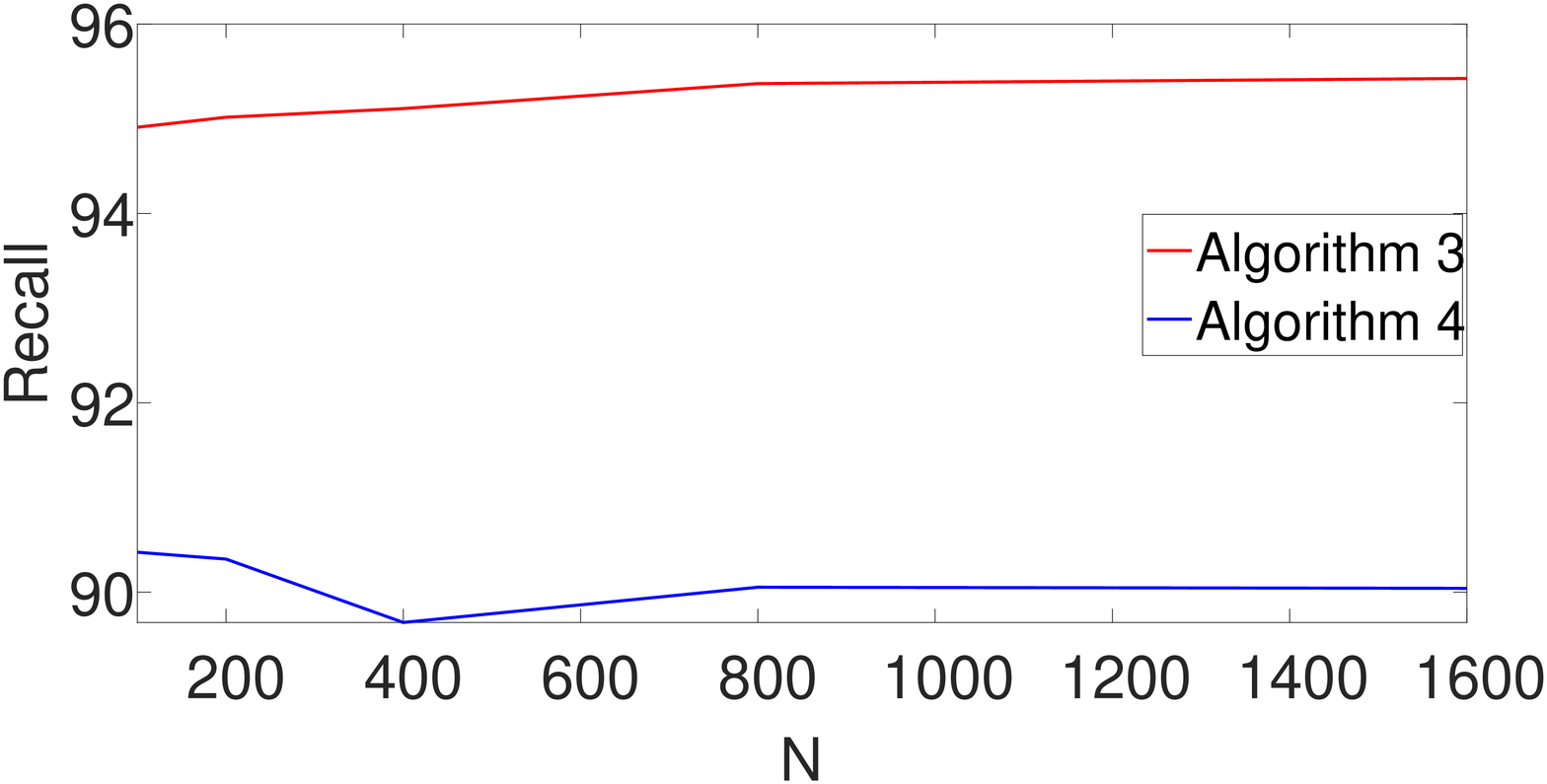}
			\caption{ Recall as a function of $N$}
		\end{subfigure}
		\hfill
		\begin{subfigure}[ht]{\columnwidth}
			\centering
			\includegraphics[width=0.7\columnwidth,bb= 20 10 1200 600,clip]{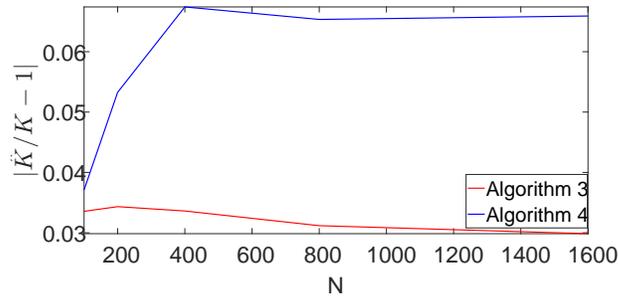}
			\caption{\label{fig:LengthIncreaseK} 
				The error in estimating the number of signal occurrences $\hat{K}$, $|\hat{K}/K-1|$, as a function of $N$}
		\end{subfigure}
		
		\caption{\label{fig:LengthIncrease} The effect of the measurement length $N$ on the estimation accuracy. %The results for Algorithm \ref{alg:Gap Statistic K find} are in red, and for Algorithm \ref{alg:Gap Statistic K find greedy} are blue.
		}
	\end{figure*}

	Next, we study the performance of the dynamic program as the length of the measurement $N$ increases. Here, we set $L=20$, and fix the density of the signals, so that  $KL/N=0.6$.
	We further assume that~$K$ is unknown. %As above, we focus on the the $F_1$-score.
	In addition to the  $F_1$-score and the recall, we also measure the accuracy of  estimating  $K$ using the measure $|\hat{K}/K-1|$. The results are presented in Figure \ref{fig:LengthIncrease}.
	
	Evidently, Algorithm~\ref{alg:Gap Statistic K find} outperforms Algorithm~\ref{alg:Gap Statistic K find greedy} in terms of  $F_1$-score, recall and the error of estimating $K$. 
	Note that our proposed scheme is not only  robust to the number of signal occurrences as $N$ grows, but it also slightly improves.
	
	\subsection{Comparison with a convex optimization approach}
	\label{sec:convex}
	
	An additional approach to detect signal occurrences is using a convex optimization framework, which was originally developed in the context of super-resolution ~\cite{candes2014towards,bendory2016robust}.
	Here, the underlying idea is to denoise the measurement using a convex program, and then apply a detection algorithm to the denoised measurement. 
	
	Here, we describe the noiseless measurement by a matrix-vector multiplication $z = {G}s$, where the $i$-th row of the circulant matrix  ${G}\in\mathbb{R}^{N\times N}$ is $x$, padded with zeros and shifted by $i$ entries, and $s\in [0,1]^N$ is a binary signal containing ones at the left-most entry of the signal occurrences and zeros otherwise. 
	The measurement is given by  $y = z + \varepsilon$, where $\varepsilon$ is i.i.d.\ white Gaussian noise with zero mean and variance of $\sigma^2$. Consequently, the detection problem to estimate the binary, sparse vector~$s$ from the measurement~$y$.  
	
	Following~\cite{bendory2016robust},  we suggest estimating $s$ by minimizing its $\ell_1$ norm subject to the  constraint $y\approx {G}s$. In addition, we relax the binary constraint to a ``box constraint,'' resulting in the following convex program:  
	\begin{equation}
		\label{KernelBendoryTV}
		\begin{split}
			&	\min_{s\in\R^{N}}||s||_1 \quad \text{subject to} \quad 	||y-{G}s||^2_2\leq\delta \\ 
			& \quad 0\leq s[i]\leq 1, \quad i=0,\ldots, N-1.
		\end{split}
	\end{equation}
	We set $\delta=1.2N\sigma^2$.
	
	We solve the convex program~\eqref{KernelBendoryTV} using CVX~\cite{grant2009cvx}, resulting a denoised measurement. Then, similarly to the procedure of Algorithm~\ref{alg:Greedy}, we chose to $K$ greatest peaks, while enforcing a separation of $L$ entries.  
	
\begin{figure}[h]
		\centering
		\includegraphics[width=0.7\columnwidth,bb= 20 10 1200 600,clip]{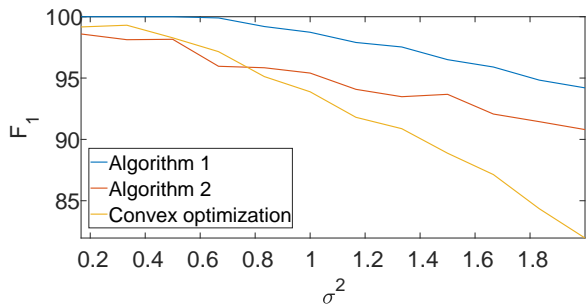}
		\caption{$F_1$-score for the arbitrary spaced model as a function of the noise level for Algorithm~\ref{alg:Ours}, Algorithm~\ref{alg:Greedy}, and the convex program~\eqref{KernelBendoryTV}}
		\label{our_vs_SR}
	\end{figure}

	Figure~\ref{our_vs_SR} compares the $F_1$-score of the convex program with the dynamic program (Algorithm~\ref{alg:Ours}) and the greedy algorithm (Algorithm~\ref{alg:Greedy}) for different noise levels. The recall is again left for the Appendix. 
	We set $L=15$, $N=75$, and $K=3$ in the arbitrary spaced setup. The dimension of the problem is relatively low because of the high computational burden of the convex approach. Evidently, both Algorithm~\ref{alg:Ours} and Algorithm~\ref{alg:Greedy} outperform the convex approach. 
	
	\section{Cryo-EM numerical experiment}
	\label{sec:Cryo Experiment} 
	In a cryo-EM experiment, biological macromolecules suspended in a liquid solution are rapidly frozen into a thin ice layer. 
	An electron beam then passes through the sample, producing a 2-D tomographic projection, called a micrograph. 
	The first step in the algorithmic pipeline is detecting the projection images in the micrograph; this process is called particle picking~\cite{heimowitz2018apple,bepler2019positive,eldar2020klt}.
	Particle picking is particularly challenging since the SNR of cryo-EM is rather low due to the absence of contrast enhancement agents and the low doses of electrons.
	The detected projection images are later used to reconstruct the 3-D structure of the sought molecule~\cite{bendory2020single,singer2020computational}. 
	The problem studied in this paper  may be viewed as a 1-D version of the cryo-EM particle picking process.
	
\begin{figure}[h]
		\centering
		\includegraphics[width=\columnwidth,bb= 20 10 1200 600,clip]{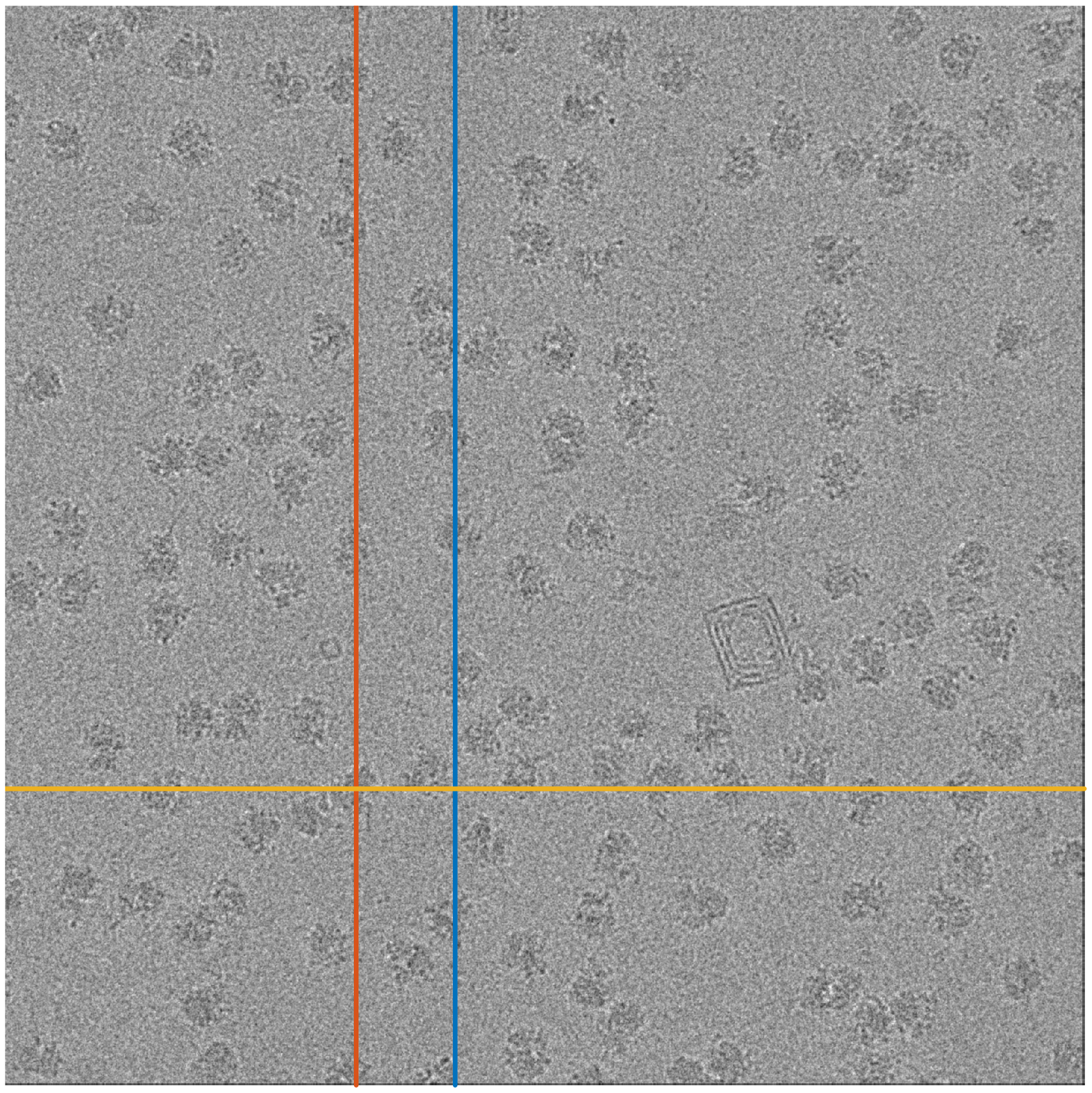}
		\caption{ A micrograph from the \texttt{EMPIAR 10028} dataset. The three marked lines are used as inputs for Algorithm~\ref{alg:Gap Statistic K find} and Algorithm~\ref{alg:Gap Statistic K find greedy}.  The red and blue lines  are columns $1324$ and $1697$, respectively, and the yellow line is row $2952$.
		}
		\label{micrograph_example}
	\end{figure}

	To test our approach, we used a micrograph that contains tomographic projections of the Plasmodium Falciparum 80S ribosome~\cite{wong2014cryo}. This data set is publicly available at the EMPIAR repository~\cite{iudin2016empiar} as \texttt{EMPIAR 10028}. 
	The micrograph is presented in Figure~\ref{micrograph_example}. 
	We arbitrarily chose  1-D stripes (columns or rows) of the micrograph, on which we can apply our 1-D detection algorithm. 
	We note that the particle projections along the   1-D stripes  are not identical, which  is a more complicated regime than the one considered in Section~\ref{sec:Experiments}.

\begin{figure}
		
		\begin{subfigure}[h]{1.2\columnwidth}
			\centering
			\includegraphics[bb=300 200 1000 400 ,clip=true,width=0.6\columnwidth]{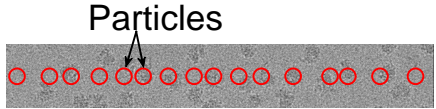}
			\caption{\label{fig:2952ourmic} Algorithm \ref{alg:Gap Statistic K find}\\}
		\end{subfigure}
		\begin{subfigure}[h]{1.2\columnwidth}
			\centering
			\includegraphics[bb=300 200 1000 400 ,clip=true,width=0.6\columnwidth]{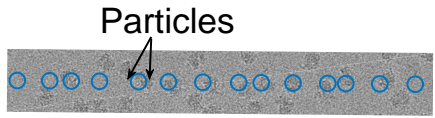}
			\caption{\label{fig:2952greedymic} Algorithm \ref{alg:Gap Statistic K find greedy} }
		\end{subfigure}
		
		\caption{\label{fig:2952} Detection using Algorithm~\ref{alg:Ours} and Algorithm~\ref{alg:Greedy}  for row 2952 (yellow line in Figure~\ref{micrograph_example}). 
			The arrows point to two particle projections, which are detected using Algorithm~\ref{alg:Gap Statistic K find} but not by Algorithm~\ref{alg:Gap Statistic K find greedy}.	
		}
	\end{figure}

\begin{figure*}
		
		\begin{subfigure}[ht]{1\columnwidth}
			\centering
			\includegraphics[trim=200 0 200 0 ,clip=true,width=0.6\columnwidth]{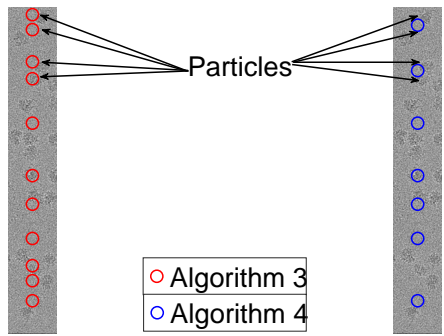}
			\caption{\label{fig:1324} Column 1324 (red line in Figure~\ref{micrograph_example})\\}
		\end{subfigure}
		\begin{subfigure}[ht]{1\columnwidth}
			\centering
			\includegraphics[trim=200 0 200 0 ,clip=true,width=0.6\columnwidth]{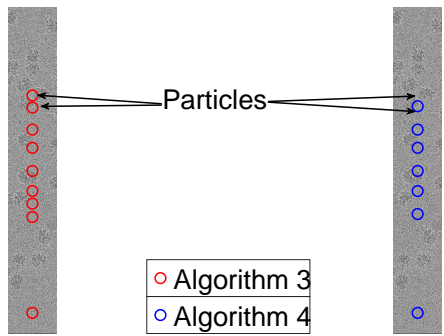}
			\caption{\label{fig:1697} Column 1697 (blue line in Figure~\ref{micrograph_example})}
		\end{subfigure}
		
		\caption{\label{fig:1324&1697} Detection using Algorithm~\ref{alg:Gap Statistic K find} and Algorithm~\ref{alg:Gap Statistic K find greedy}. 	The arrows point to  particle projections, which are detected by Algorithm~\ref{alg:Gap Statistic K find}, while Algorithms~\ref{alg:Gap Statistic K find greedy} fails. }
	\end{figure*}

\begin{figure*}
		
		\begin{subfigure}[h]{1\columnwidth}
			\centering
			\includegraphics[width=0.65\columnwidth]{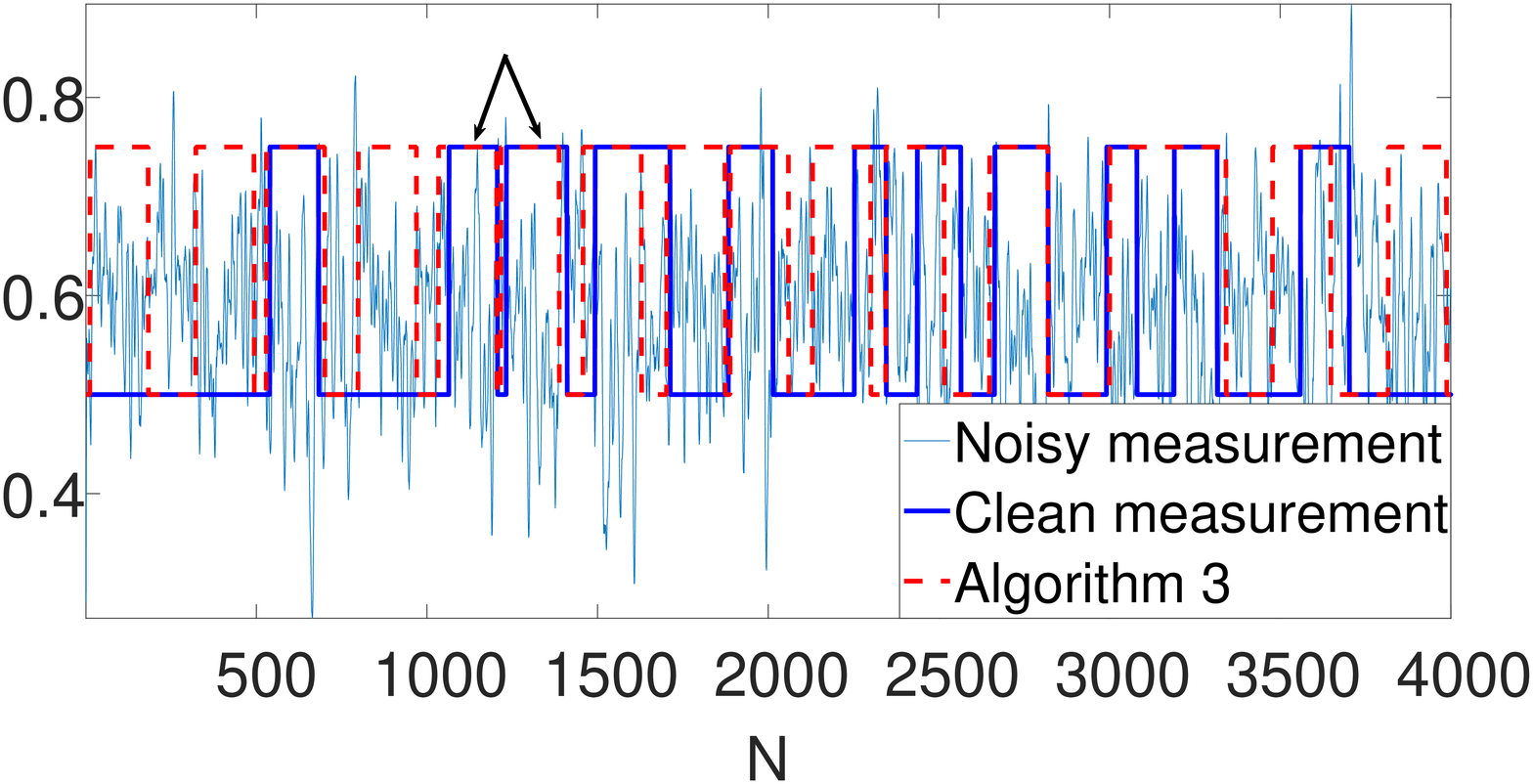}
			\caption{ Algorithm \ref{alg:Gap Statistic K find}\\}
		\end{subfigure}
		\begin{subfigure}[h]{1\columnwidth}
			\centering
			\includegraphics[width=0.65\columnwidth]{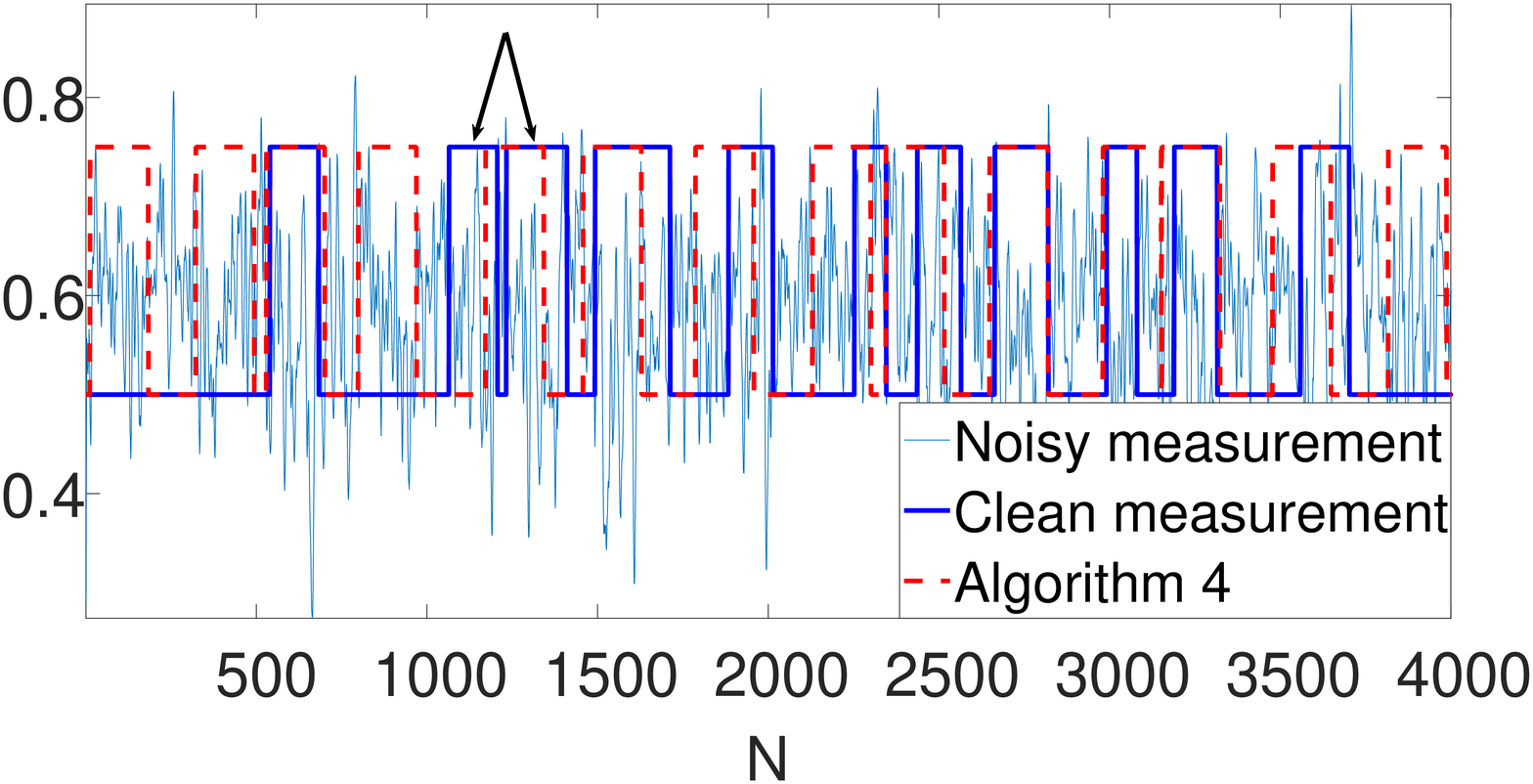}
			\caption{ Algorithm \ref{alg:Gap Statistic K find greedy} }
		\end{subfigure}
		
		\caption{\label{fig:2952_1D} Detection using Algorithm~\ref{alg:Ours} and Algorithm~\ref{alg:Greedy}  for row 2952 (yellow line in Figure~\ref{micrograph_example}). 
			The arrows point to two particle projections, which are detected by Algorithm~\ref{alg:Gap Statistic K find} and not by Algorithm~\ref{alg:Gap Statistic K find greedy}.	
		}
	\end{figure*}

	As a prepossessing step, we whiten the noise: a standard step in many cryo-EM algorithmic pipelines. 
	This is done in the following manner. First, we manually find a region in the measurement with no signal. Using this ``noise only'' data, we  approximate the power spectral density of the noise. Then, we multiple the entire measurement by the inverse of the approximated power spectrum.  We are now ready to apply Algorithm~\ref{alg:Gap Statistic K find} and Algorithm~\ref{alg:Gap Statistic K find greedy} to 1-D measurements, after whitening. 
	We assumed that the shape of the signal is a square pulse whose length is chosen manually.
	To evaluate the results, we manually tagged the true locations of the particles (namely, signal occurrences). Figure~\ref{fig:2952} and Figure~\ref{fig:1324&1697} illustrate the results.  
	While both algorithms are fairly similar in the more sparse environments, the dynamic program approach succeeds in identifying densely packed particles (highlighted with arrows), while the greedy method fails. This indicates that extension of our scheme to 2-D images may be helpful to locating densely packed particle images in cryo-EM data sets. In addition, we illustrate 1-D projections of our results in Figure \ref{fig:2952_1D}. As we can see, our proposed scheme successfully detects the particles while the greedy algorithm demonstrates inferior results. 
	%\mordechai{A 1-D projection of the results in Figure ~\ref{fig:2952} and Figure~\ref{fig:1324&1697} appear in Figure ~\ref{fig:2952_1D},Figure ~\ref{fig:1324_1D} and Figure ~\ref{fig:1697_1D}}.
	
	\section{Discussion} \label{sec:conclusions}
	
	This papers introduces a novel scheme for signal detection based on a dynamic program that  maximize a constrained likelihood function. We apply the gap statistic principle to estimate the number of signal occurrences, and provide an end-to-end solution to the problem. We demonstrate our proposed method in a series of experiments. Our suggested scheme demonstrates improved performance over popular alternatives in dense environments, while attaining similar results in sparse regimes. This makes it a robust approach in many practical setups.  
	
	Our work is motivated by the cryo-EM technology. Typically, particle pickers are based on cross correlating the micrograph with different templates. This approach performs well in cases where the particles are well separated but fails in dense regimes. 
	We show that by imposing a separation constraint, we improve upon  currently known schemes in the 1-D regime. 
	This motivated our future work,  generalizing our results to 2-D images, and provide an efficient solution to the cryo-EM particle picking problem. 
	
	\section*{Acknowledgment}
	This work was supported  by the Tel Aviv University Center for Artificial Intelligence and Data Science.
	T.B.  is also partially supported by the  NSF-BSF award 2019752, the BSF grant no. 2020159, and the ISF grant no. 1924/21. 
	A.P. is  partially supported by the ISF grant no. 963/21.

\bibliographystyle{plain}
%\bibliographystyle{IEEEtran}
%\bibliography{sigproc}

%\input{Signal_detection_with_dynamic_programming.bbl}

\end{document}